# EL ELIXIR DE LA ENERGÍA ETERNA

José Manuel Quesada Molina
Departamento de Física Atómica, Molecular y Nuclear
Universidad de Sevilla

**Abstract**

The recent announcement of a purported breakthrough result in inertial nuclear fusion at NIF (Lawrence Livermore Laboratory, USA) has aroused a tide of media and public interest. The excitement has been generalized to the whole field of research in fusion energy with, in its wake, announcements of an imminent advent of the cure for the energetic crisis and the aggravating influence in the climate change associated to the fossil fuels. This opinion article is intended to show that such expectations are not founded on sound scientific bases and that there is a long way until the practical production of electricity from nuclear fusion is achieved, if ever.

**Resumen**

El reciente anuncio de un supuestamente trascendental resultado en fusión nuclear inercial en NIF (Lawrence Livermore Laboratory, EEUU de Norteamérica) ha desatado un enorme interés en el público y los medios de comunicación. El entusiasmo se ha trasladado a todo el campo de la investigación en fusión para la producción de energía con, a su estela, anuncios de la llegada inminente de la solución a la crisis energética y al efecto agravante del cambio climático asociado a los combustibles fósiles. Este artículo de opinión pretende poner de manifiesto que tales expectativas no están fundadas en bases científicas sólidas y que hay un largo camino por recorrer hasta que se logre, a niveles prácticos, la producción de electricidad a partir de la fusión nuclear, si se consigue alguna vez.

## Introducción

Recientemente se anunció con extraordinario aparato mediático un nuevo hito alcanzado en la fusión nuclear, muy oportunamente publicitado en el contexto actual de crisis energética. En principio no hay nada que objetar a ello: en la comunidad científica es bien conocida la necesidad de financiación que tienen los grupos de investigación, y no digamos los grandes laboratorios como el Lawrence Livermore Laboratory, que deben anunciar sus logros científicos a fin de despertar el interés de la opinión pública, que busca soluciones tangibles a sus problemas inmediatos, y, con ello, captar la atención de las instancias políticas, siempre ávidas de réditos demoscópicos.

No obstante, opino que se está vertiendo con demasiada frecuencia información sesgada que induce a confusión. Comencemos con el detonante del presente texto, la noticia que ha desencadenado el frenesí mediático: el DOE (*Department of Energy*) norteamericano anunció con gran parafernalia mediática el pasado 13 de diciembre que en la instalación NIF (*National Ignition Facility*) del Lawrence Livermore Laboratory se acababa de conseguir superar el *breakeven* en un experimento de fusión, lo cual simplemente consiste en conseguir más energía que la suministrada. Suena muy bien y prometedor, pero conviene puntualizar. En primer lugar, esto se ha conseguido mediante la compresión con 192 láseres de altísima potencia sincronizados en un brevísimo pulso del orden de varios nanosegundos[1] (la mayor concentración energética mediante láser jamás conseguida) de una diminuta cápsula de diamante conteniendo un *pellet* ultracongelado con dos isótopos del hidrógeno, deuterio y tritio (DT)[2]. Ello es un mecanismo completamente diferente al confinamiento magnético que se utiliza en los llamados *tokamaks*[3], donde el confinamiento se consigue mediante campos magnéticos con geometría toroidal (*donut*). En la actualidad el diseño tipo *tokamak* es el más ampliamente utilizado, particularmente en Europa, donde bajo estas premisas desde comienzos de los años 90 se realizan experimentos en el JET (*Joint European Torus*) en Gran Bretaña, con financiación de la Unión Europea (al menos era sí hasta el *Brexit*, ya que actualmente está en fase terminal), y actualmente está en construcción el ITER (acrónimo de International Thermonuclear Experimental Reactor) en Cadarache, Francia. Por lo tanto, el logro alcanzado en el NIF es muy difícilmente extrapolable a la apuesta científico-técnica mayoritaria en curso, que es el confinamiento magnético; es más, dicha vinculación se me escapa por completo. Por lo tanto, considero que ésta es una matización clave, que se debe realizar claramente desde el principio: ambos mecanismos (compresión inercial – mediante láser u otro tipo de haces de partículas - y confinamiento magnético), aunque comparten finalidad, difieren esencialmente y los pretendidos logros en uno no son trasladables al otro.

**¿Es todo esto como se anuncia y promete?**

A raíz de la noticia, y aprovechando su tirón mediático, he podido leer nuevamente (para mi gran sorpresa) algo que lleva repitiéndose desde los orígenes del desarrollo de la fusión nuclear: se trataría de reproducir en la Tierra el proceso que tiene lugar en el Sol y permite la vida; es más, se le llega a poner fecha: un reactor de fusión conectado a la red eléctrica en el próximo decenio. Lo primero no es cierto en sentido estricto y lo segundo, que contradice los ya de por sí optimismas y largos plazos de ITER-DEMO, como se verá a continuación, no me lo creo. El tiempo dirá y la hemeroteca lo reflejará.

En el Sol se *quema* hidrógeno, un recurso inagotable a nuestra escala, para producir partículas alfa, que son estables, es decir sin producir residuos radiactivos[4]. Todo extraordinariamente prometedor, salvo por un detalle: no es posible realizarlo en la Tierra, ya que entra dentro del dominio de la ciencia ficción. El motivo: la probabilidad de que dos núcleos de hidrógeno (es decir, dos protones) superen su repulsión eléctrica mutua y se fundan es tan pequeña que ni siquiera ha podido medirse experimentalmente en un laboratorio. En el Sol ello se produce debido a las *monstruosas* (a escala terrestre) densidades de masa que se alcanzan en su interior por la atracción gravitatoria de su enorme masa (también comparada con la de la Tierra); pero en la Tierra no son alcanzables tales densidades[5]. Este es el motivo de que haya que recurrir a otras mezclas de *fusionantes* : deuterio-deuterio (DD), la ya mencionada deuterio-tritio (DT), etc.. Es decir, lo que se pretende realizar en la Tierra es parecido a lo que ocurre en el Sol, pero no es lo mismo; en ambos casos hay un mecanismo común, pero el combustible es diferente. En particular, la única que se vislumbra con posibilidades de permitir la fusión para producir energía eléctrica es la combinación DT[6], que es la adoptada en todos los proyectos vigentes que pretenden conducir a ese objetivo.

El deuterio es abundante (constituye una pequeña fracción del hidrógeno natural) y estable. Pero el tritio no es ninguna de las dos cosas: es radiactivo y, por lo tanto, no existe naturalmente; es decir, hay que producirlo. Esto cambia bastante el panorama de supuestas bondades del combustible (*casi infinito*, según se anuncia): el tritio, como isótopo del hidrógeno, se comporta químicamente (y, por lo tanto, biológicamente) exactamente igual que el hidrógeno *normal* ; es decir, dado el papel central del hidrógeno en el ciclo de la vida, el tritio se incorpora al mismo sin que haya forma de separarlo químicamente, porque es hidrógeno (aunque radiactivo). Su vida media de 12.33 años hace que en ese tiempo su cantidad se reduzca a la mitad, pero en un reactor de fusión ha de producirse continuamente, para lo cual está planeado, en una fase posterior de ITER, colocar en la *manta* exterior (por la que circulará el agua de refrigeración) litio. Los problemas derivados de la actual escasez de litio, debida a la creciente demanda de baterías y causante de que su precio se haya disparado recientemente, y la contaminación asociada a su minería intensiva son conocidos[7]. Además, el litio natural contiene un 92.41% del isótopo Li7 y un 7.59% del Li6, que es el isótopo fértil para producir tritio con neutrones ya *termalizados*[8], como se pretende. Será preciso aumentar la proporción de Li6, proceso que no es fácil ni barato (como tampoco lo es en el caso del uranio), ya que no se puede hacer por procedimientos puramente químicos. La situación planteada en ITER es, cuando menos, paradójica: incialmente se propone utilizar tritio producido en reactores nucleares de fisión (tecnología ya probada, que pretende superar) y en una fase posterior comprobar la factibilidad de la regeneración del tritio a partir del litio (que actualmente no es barato ni limpio en su orígen) en la tasa suficiente para mantener la reacción[9]. Todos estos aspectos deben incluirse en el balance global, que sistemáticamente se omite. La provisión de litio es otro de los aspectos que quedan supeditados a futuros avances. Nuevamente nos movemos en el campo de las expectativas.

Las consecuencias de la infiltración y fuga del tritio a través de las paredes del reactor a las enormes temperaturas a las que se pretende que funcione se conocen sólo parcialmente, ya que los valores que se manejan se basan en extrapolaciones. Por lo tanto el combustible previsto no es casi infinito, ni es limpio ni seguro ni barato. El tritio es tan problemático[10] que en JET se ha trabajado en la medida de lo posible sólo con hidrógeno o sólo con deuterio, *extrapolándose* las tasas de reacción a la mezcla DT. De este modo se obtiene el factor Q (o eficiencia energética, que es el cociente entre la potencia conseguida y la consumida; hablaremos de él más adelante) *extrapolado*, con el llamativo resultado de que cuando se realizó el experimento con la mezcla DT el valor de Q obtenido fue aproximadamente la mitad; lo cual muestra algo bien conocido en física e ingeniería, que es el riesgo de las extrapolaciones, a la vez que pone en evidencia la problemática asociada al uso del tritio. Lo anterior añade otra incógnita más a la pretendida *limpieza radiológica* de la fusión para producir energía eléctrica a partir de combustible *limpio, barato, accesible e inagotable,* según otro lugar común en muchas declaraciones leídas en la prensa: "porque se extrae del agua del mar".

**Jugando con las definiciones**

Otro aspecto a destacar de la citada noticia tiene que ver con el ya mencionado logro del *breakeven* en el NIF. Este consistió en alcanzar un factor Q de ganancia energética de fusión de 1.54, es decir que se obtuvo 1.54 veces más energía que la se invirtió . Al margen de otras consideraciones en las que entraré más adelante, resulta llamativo (por expresarlo suavemente) el cambio de definición que ha conducido a este anunciado *éxito*. El NIF cambió hace algunos años la definición del Q para colocar en el denominador (potencia que hay que suministrar a los láseres para comprimir y calentar el plasma) sólo la fracción que éstos devuelven en forma de radiación ultravioleta para comprimir y calentar, es decir sólo la fracción aprovechable. Teniendo en cuenta que la eficiencia de los láseres es muy baja (en torno al 1%), en rigor hay que dividir por toda la energía invertida, es decir dividir el aunciado factor de ganancia $Q = 1.54$ por 100, con lo cual se está aún muy lejos de recuperar lo invertido. Muy lejos. Obviamente, esta redefinición unilateral del factor $Q$ por parte del NIF recibió severas críticas[11], pero el hecho de que no se haya reflejado en las noticias esta *matización* (¡de un factor 100!) por parte de sus voceros (o al menos yo no la he encontrado) permite hacerse una idea del poder del *lobby* que hay detrás.

Toda la discusión anterior se ha realizado omitiendo un detalle adicional que considero fundamental para tener una visión clara de la situación real: no toda la energía liberada en la fusión (el numerador del factor Q) es aprovechable para producir calor y, con ello, la energía eléctrica que se pretende obtener en última instancia. En la fusión DT el 80% de la energía producida (17.6

MeV[12]) se la llevan los neutrones muy rápidos (14.1 MeV) en forma de energía cinética, siendo las partículas alfa (3.5 MeV), que se llevan el 20% restante, las responsables de la mayor parte del calentamiento[13]. Por el contrario, en una fisión del combustible típico de las centrales nucleares de fisión (U235), sólo en torno al 5% de la energía liberada (unos 200 MeV) se la llevan los neutrones (bastante más lentos que en el caso de la fusión DT, siendo su energía promedio 2 MeV), mientras que el resto corresponde a los fragmentos de fisión, núcleos de tamaño medio muy cargados eléctricamente, que son los responsables del calentamiento de las barras de combustible, que a su vez calientan el refrigerante (normalmente agua) encargado de transportarlo. Por lo tanto, incluso sin la redefinición del NIF, el factor Q dista de ser una medida realista de la rentabilidad energética del proceso de fusión, ya que no sólo una fracción minoritaria de la energía liberada en la fusión es aprovechable para producir calor, que es lo que interesa.

En la misma línea de información sesgada por parte de los gabinetes de comunicación, ITER hizo oficialmente pública una información que claramente conduce a error de interpretación. Concretamente, se afirmaba que ITER será capaz de producir 500 MW[14] de potencia a partir de 50 MW de potencia suministrada. De ahí se infería lógicamente que esos 50 MW suministrados se referían a la potencia total eléctrica invertida, no a la calorífica finalmente suministrada al plasma (es decir, igual que en el caso de NIF con los láseres). Ante las críticas recibidas[15], tuvieron que rectificar y ahora se especifica *heating input power*, es decir, potencia de calentamiento aportada al plasma. Debido a la la eficiencia del proceso de conversión a partir de la energía eléctrica que es preciso aportar (siempre menor que la unidad, usualmente mucho menor, como ocurre en el caso del NIF, donde es del orden de una centésima), ésta última es muy superior, estimándose que son necesarios entre 200 y 300 MW eléctricos para iniciar y mantener la fusión[16]. Ya que nos movemos en el ámbito de las estimaciones, no encuentro motivo a priori valorar unas más que otras: sólo la experiencia dará su veredicto. Además, los 500 MW producidos son totales en forma de energía cinética de los productos, de los cuales, como ya se ha indicado, aproximadamente el 80% corresponde a los neutrones muy rápidos, que son mucho menos eficientes produciendo calor (sólo son capaces de transferir una parte del mismo al medio antes de escapar), que, al no ser utilizable para mantener la temperatura del plasma, ITER propone[17] aprovechar calentando el agua del circuito refrigerante de la *manta* que envolverá la cámara de vacío. Toda la *responsabilidad* del mantenimiento de la temperatura del plasma recaerá sobre las partículas alfa, que depositarán directamente en el mismo su energía (que, recordemos, es sólo el 20 % de la energía en cada fusión). La regeneración del tritio (caro, escaso y que debe producirse en reactores nucleares de fisión, principalmente) necesario para mantener la fusión se pospone para para una fase posterior de ITER, donde se experimentará con la capacidad del isótopo minoritario Li6 del litio natural (que debe ser enriquecido para ello) para producirlo en la suficiente cantidad para mantener la reacción[18]. Nuevamente nos movemos en el campo de las expectativas, sólo la experimentación demostrará la viabilidad de la propuesta.

**Un poco de historia**

Las radicales diferencias entre los procesos de fusión y fisión nuclear son la causa de que entre la primera reacción nuclear explosiva en cadena (*Trinity* , 16 de julio de 1945 en Alamogordo, Nuevo México, EEUU) y la primera producción comercial de energía eléctrica mediante la fisión controlada[19] (18 de diciembre 1957, en Shippingport, Pennsylvania, EEUU) transcurriesen solamente 12 años, mientras que tras la primera explosión termonuclear[20] (*Ivy Mike*, 1 de noviembre de 1952 en el atolón Enewetak, en las Islas Marshall) aún no se ha conseguido *domesticar* la fusión para mantenerla bajo control y producir energía aprovechable.

En el caso de la fisión se pretende (y consigue desde el año 1942) mantener controlada una reacción en cadena, donde los garantes de esa continuidad son los neutrones producidos en cada reacción. En el caso de la fusión los neutrones no juegan ningún papel en mantenimiento de la misma, sino que el agente garante de la reacción en cadena es el calor producido, que se debe traducir en temperatura (manteniendo la densidad). Cuando no se pretende el control de la misma sino todo lo contrario (bombas), ello se hace por *fuerza bruta* (nunca mejor dicho) recurriendo a la extraordinaria presión de radiación originada por el *fulminante* de fisión[21], sin que ésta escape antes de conseguir instantáneamente su objetivo (todo ocurre en unos pocos microsegundos[22]). En cambio, para mantener la reacción de fusión en un reactor no se puede, obviamente, recurrir a ese mecanismo explosivo y se debe conseguir que el calor generado por las reacciones de fusión se recicle sin escapar para mantener la temperatura, al tiempo que la densidad se mantenga temporalmente. Una empresa formidable, que aún está por conseguirse.

### ¿Por qué se persigue conseguir tan elevadas densidades y temperaturas en un futuro reactor nuclear de fusión?

Porque es preciso conseguir que núcleos atómicos ligeros superen la repulsión debida a su carga eléctrica y se fundan en un núcleo mayor y alguna otra partícula emergente; y además que lo hagan en la tasa (velocidad a la que se producen las reaccciones) suficiente. La clave radica en que la suma de las masas de los productos de la reacción es ligeramente inferior a la masa de los reaccionantes, convirtiéndose esa diferencia de masa $m$ en energía $E$, según la archiconocida fórmula de Einstein $E=m c^2$, donde $c$ es la velocidad de la luz. Este mecanismo es el opuesto al de la fisión nuclear, aunque la consecuencia es la misma: conversión de masa en energía. En la fisión un núcleo pesado captura un neutrón y se rompe en dos fragmentos de aproximadamente la mitad de su masa y varios neutrones. Aquí tenemos por tanto una gran diferencia cualitativa: a diferencia de la fusión, en la fisión el agente desencadenante (el neutrón, que como su nombre indica, carece de carga eléctrica) no tiene que superar en primer lugar la repulsión eléctrica por parte del núcleo (donde hay protones y neutrones que, en todo caso, lo atraen por la llamada interacción nuclear o fuerte, que es de corto alcance). Por ello en la fisión, si se dan las circunstancias adecuadas (características del núcleo progenitor y energía del neutrón incidente) el núcleo compuesto resultante, que se forma en un estado excitado, se rompe *espontáneamente* en busca de una mayor estabilidad del sistema, es decir *fisiona.* Nos encontramos ante una situación radicalmente diferente a la de la fusión, donde los dos intervinientes han de superar su repulsión mutua (ambos están cargados positivamente), lo cual implica enormes temperaturas para conseguirlo[23]. Además la densidad ha de ser altísima para que la tasa de reacción sea la suficiente, como comentaré a continuación.

**La tasa de reacción es la clave**

La tasa de una reacción nuclear (es decir, el número de reacciones por unidad de tiempo) es proporcional a la densidad de blancos[24], al flujo de proyectiles que los bombardean y a la probabilidad de que la reacción se produzca una vez que proyectil y blanco colisionan. Esta última cantidad es a su vez proporcional a una magnitud llamada sección eficaz[25], que viene determinada por la estructura nuclear intrínseca de cada pareja proyectil-blanco y en la cual nuestro margen de maniobra está limitado a la velocidad relativa, es decir, a la temperatura. Por lo tanto, para cada pareja de proyectil y blanco reaccionantes (fusionantes o fisionantes), conseguir una tasa de reacción suficiente exige unos valores adecuados de densidad y temperatura.

La tasa de reacción es la clave para producir energía aprovechable, porque las reacciones de fusión se producen rutinariamente en laboratorio mediante el uso de aceleradores (controladas, pero no automantenidas ya que exigen aporte continuo de energía). Una de las fuentes habituales de neutrones es la llamada DT (deuterio-tritio), la misma mezcla prevista en ITER, en la cual mediante un acelerador se bombardea con deuterones un blanco de tritio gaseoso, en cada una de cuyas reacciones de fusión, como se ha indicado anteriormente, se liberan unos 17.6 MeV de energía, que se reparten entre una partícula alfa, que se lleva aproximadamente el 20% de dicha energía, y un neutrón, que se lleva el 80% restante. Pero la producción energética en forma de calor (como se ha visto, debido mayoritariamente a la energía que transportan las partículas alfa) es ínfima debido a los valores de las tasas de reacción implicadas. Es decir, esta fusión DT ( y lo mismo se puede decir de las fuentes de neutrones DD) no sirve para producir energía eléctrica aprovechable, su finalidad es producir neutrones rápidos.

**¿Por qué es tan difícil conseguir la fusión nuclear automantenida?**

En el caso de la fisión, en la que se basan las centrales nucleares actuales, la densidad de núcleos está fijada por ser el combustible un medio sólido (aunque puede ser líquido, que para el caso es lo mismo) y la sección eficaz (es decir, la probabilidad de que se produzca la reacción) se hace enorme en los núcleos fisionables para una energía adecuada de los neutrones (recordemos que los neutrones no tienen carga eléctrica, por lo que se *cuelan* sin obstáculo en los núcleos blanco, y que la sección eficaz varía con su energía – es decir con la temperatura del medio que los termaliza, es decir que los frena- debido a los detalles de la estructura nuclear). Por este motivo, para mantener bajo control la tasa de reacciones nucleares basta con controlar con precisión la población neutrónica. Con ello se consigue una tasa de reacción que libera la cantidad de calor suficiente para ser transformado comercialmente en energía eléctrica.

Por el contrario, en el caso de la fusión (la que nos anuncian como fuente de energía limpia e inagotable del futuro) la sección eficaz es extraordinariamente pequeña comparada con la de fisión. Concretamente, en la fusión DT la sección eficaz a la temperatura prevista de unos 150 millones de grados [26] es aproximadamente una diezmilésima de la sección eficaz de fisión de un núcleo de U235 bombardeado por neutrones termalizados (es decir con energía óptima para fisionar eficazmente este isótopo) en un reactor convencional refrigerado por agua ligera a presión (LWR-PWR), que opera a unos 350ºC. Conseguir la temperatura necesaria en el plasma de fusión (que es la *sopa* de núcleos y electrones en que se transforma la materia esas temperaturas), es ya en sí misma una empresa formidable, pero a eso hay que añadir la necesidad de alcanzar una densidad suficiente de dicho plasma y, además, que ambos parámetros se mantengan durante el tiempo suficiente para mantener una tasa de reacción que la haga utilizable para producir energía. En las bombas *de fusión* (empleadas con fines bélicos) se utilizan una o varias bombas de fisión para conseguir simultáneamente los objetivos anteriores (compresión y calentamiento), y en ellas, obviamente, ni el control ni el mantenimiento temporal son necesarios, sino todo lo contrario, desafortunadamente. Pero evidentemente este mecanismo está excluido para aplicaciones pacíficas, de forma que el objetivo de mantener el plasma en esas condiciones sólo se plantea de forma pulsada, ya sea mediante compresión con láseres o confinamiento magnético (con la que se pretende llegar a varios centenares de segundos). Ello explica el ya mencionado hecho de que desde *Ivy Mike* en 1952 hayan transcurrido 70 años sin alcanzar la fusión controlada para la producción comercial de energía eléctrica, siendo las predicciones más optimistas de unos 30, 40, ¿50? años adicionales para conseguirlo. Porque ITER, cuando funcione, está destinado a ser *la prueba de concepto científica* de la fusión controlada y automantenida para producir más energía energía cinética de los productos (la que se denomina *fusion power,* potencia de fusión, definición que se presta a confusión porque es mucho mayor que la calorífica, que es la aprovechable parcialmente para producir energía eléctrica, como se ha visto) que la potencia calorífica *efectiva* en el plasma usada para calentarlo (la que se denomina *input heating power*[27], potencia entrante de calentamiento, también una definición que se presta a confusión, porque es mucho menor que la total que se debe suministrar externamente para confinar y calentar, como también se ha visto). Habrá que esperar a DEMO (como su nombre indica) para la *prueba de concepto de ingeniería*, que demuestre que es posible verter energía neta a la red eléctrica, es decir producir más energía eléctrica que la energía eléctrica invertida. Y mientras tanto debe continuarse investigando exhaustivamente en los efectos que el extraordinario bombardeo con neutrones de alta energía a semejantes temperaturas induce en las propiedades de los materiales estructurales del reactor, en particular la fragilización, aparición de fallas y deformaciones [28]. Y tras todo ello, si se alcanza ese punto (cosa que en el mejor de los casos podrán ver nuestros nietos o bisnietos, porque ninguno de nosotros tendrá la oportunidad de sonreir consultando la hemeroteca) habrá de demostrarse la viabilidad económica de esta fuente de energía, que dados los enormes costes de desarrollo y su descomunal consumo energético previo acumulado en forma de consumo de combustibles fósiles y energía eléctrica de origen nuclear de fisión (hay estudios al respecto) dista mucho de estar claro.

**El balance energético**

En principio, un argumento que parece favorecer a la fusión nuclear frente a la fisión es la energía específica (o energía por unidad de masa atómica). Vamos a explicarlo. Una característica de los núcleos atómicos, aunque no exclusiva, ya que lo siguiente es aplicable a cualquier sistema regido por las leyes de la Física Cuántica (es decir a todos), es que su masa es menor que la suma de las masas de sus constituyentes por separado. Esa diferencia, traducida en energía por la fórmula de Einstein, es lo que se conoce como energía de ligadura. Por lo tanto si en una reacción nuclear pasamos de una situación con menos ligadura (más masa) a otra de más ligadura (menor masa), la diferencia se transforma en energía cinética de los productos y radiación. Y esa es la energía que, en forma de calor, se utiliza (en un reactor nuclear de fisión) o debería algún día poder utilizarse (en un reactor nuclear de fusión) para producir energía eléctrica. Si colocamos los isótopos conocidos (es decir tipos de núcleos atómicos) en orden creciente con sus masas, comenzando en el hidrógeno, encontramos que la ligadura por nucleón va aumentando en promedio hasta el Fe56 (núcleo de hierro con 26 protones y 30 neutrones, donde alcanza casi 9 MeV por nucleón); a partir de ese punto comienza a disminuir suavemente hasta el final de la tabla, donde llega a unos 7.5 MeV por nucleón en la región de isótopos que nos interesa (la llamada zona de los actínidos). Ello quiere decir que, cuando dos isótopos ligeros (por debajo de la masa del Fe56) se fusionan, el resultado está más ligado en general, es decir tiene menos masa, y esa pérdida de masa se transforma en energía. Por ejemplo en la típica reaccción DT la ligadura del deuterón son unos 2.2 MeV, es decir aproximadamente 1.1 MeV por nucleón ; la ligadura del tritio son unos 8.5 MeV, es decir unos 2.8 MeV por nucleón; la ligadura de la partícula alfa son unos 28.3 MeV, es decir aproximadamente 7.1 MeV por nucleón. Por lo tanto, se pasa de una ligadura incial en el sistema DT de aproximadamente 10.7 MeV a los 28.3 MeV de la partícula alfa, es decir hay una ganancia de energía ligadura de unos 3.5 MeV por nucleón inicial (~17.6/5) eV. La situación opuesta se presenta en el otro extremo de la tabla de isótopos cuando un núcleo pesado, por ejemplo el U235, captura un neutrón y fisiona. La energía de ligadura del U235 son unos 1786.7 MeV, es decir aproximadamente unos 7.6 MeV por nucleón, y la de dos típicos productos de fisión (recordemos que es un proceso probabilístico) está próxima a la máxima del Fe56, digamos que en torno a los 8.5 MeV por nucleón; por lo tanto se ha ganado aproximadamente 0.9 MeV por nucleón en energía de ligadura. Multiplicando esta cantidad por los 236 nucleones del núcleo compuesto inicial (el de U235 más el neutrón absorbido) resultan unos 212 MeV de energía liberada en una fisión típica, cantidad bastante próxima a los valores medidos experimentalmente. Desde este punto de vista, en principio parecería que la fusión DT es más interesante, ya que la ganancia neta de ligadura por nucleón del combustible es casi 4 veces mayor (lo cual se traduce en una energía específica unas 4 veces mayor del combustible DT respecto del U235), pero hay que considerar que la mayor parte de esa energía cinética corresponde a los neutrones rápidos (el ya mencionado 80% típicamente en el caso de la fusión DT, frente al aproximadamente 5% en el caso de la fisión del U235, donde, además, en promedio son mucho más lentos), que es muy poco aprovechable para producir calor, es decir, en última instancia energía eléctrica. En resumidas cuentas, un reactor de fusión es una magnífica fuente de neutrones muy energéticos, otro asunto muy diferente es cómo aprovechar la energía producida (de la que esos neutrones se llevan lel 80%), ya que sólo una pequeña parte de ella se podrá transformar en calor.

**Algunas consideraciones adicionales al balance energético**

En la sección anterior he preferido mantener la discusión en torno a la energía específica, que es un argumento que con frecuencia se emplea en favor de la fusión, ya que de él se deduce, en mi opinión erróneamente, que la fusión es más eficiente que la fisión en términos energéticos. Sin embargo, en la práctica, lo que tiene relevancia para la producción global de energía (que, no olvidemos, es el objetivo último perseguido) es la energía liberada en cada reacción, que, como acabamos de ver, supone 17.6 MeV en la fusión DT frente a unos 200 MeV en la fisión del U235. El motivo es que para obtener la potencia generada ésta es la cantidad que hay que multiplicar por el número de reacciones por unidad de tiempo, es decir, por la tasa de reacción, que, como también se ha indicado anteriormente, es la clave para obtener el objetivo pretendido. A ésta última contribuyen multiplicativamente la densidad de núcleos blanco, la sección eficaz y el flujo de partículas incidentes (número de partículas por unidad de área y de tiempo).

En cuanto a la densidad, un cálculo trivial resulta en que la densidad de núcleos de U235 en el óxido de uranio ($UO_2$, densidad=10.97 g $cm^{-3}$) enriquecido al 3%, que es combustible de un reactor nuclear típico LWR-PWR, es aproximadamente $10^{27}$ $m^{-3}$, mientras que la densidad que se planea alcanzar en ITER[29] es aproximadamente $10^{20}$ $m^{-3}$; es decir, hay 7 órdenes de magnitud (diez millones) de diferencia.

La sección eficaz (que, como ya se ha indicado, es una medida de la probabilidad de que se produzca la reaccción nuclear cuando los reaccionantes "se encuentran") es 4 órdenes de magnitud (diez mil veces) superior en el caso de la fisión de U235 a los típicos 350 º centígrados en un reactor de fisión LWR-PWR frente a la de la fusión DT a los pretendidos 150 millones de grados en ITER.

En cuanto al flujo de partículas incidentes, en el caso de un reactor de fisión LWR-PWR es del orden de $10^{13}$ neutrones $cm^{-2}$ $s^{-1}$. Se puede realizar una estimación para el caso del plasma de fusión DT en ITER, teniendo en cuenta que dicho flujo es el producto de la densidad por la velocidad; ésta última se puede estimar mediante un sencillo cálculo no relativista para deuterones a una temperatura de 150 millones de grados en unos $10^7$ m/s, lo cual se traduce en un flujo del orden de $10^{23}$ deuterones $cm^{-2}$ $s^{-1}$ (el considerar nos núcleos de tritio como las partículas incidentes no cambia el orden de magnitid de las estimaciones). Es decir el flujo incidente en el caso de fusión DT en ITER será unos 10 órdenes de magnitud superior al de neutrones en el caso de la fisión del U235 en un reactor típico.

Multiplicando los tres factores anteriores (densidad, sección eficaz y flujo) llegamos a la conclusión de que la tasa de reacción en la fusión DT de ITER debería ser unas 10 veces inferior a la de la fisión del U235 en reactor típico. Teniendo en cuenta que en cada reacción de fisión se liberan unas 10 veces la energía de una fusión DT, resulta que la tasa de producción de energía (es decir, la potencia) total producida en forma de energía cinética de los productos sería, según estas someras estimaciones, unos dos órdenes de magnitud (un factor 100) superior en la fisión del

U235 en reactor *convencional* respecto a la fusión DT en ITER. Pero, como ya se ha indicado, esto son someras estimaciones, por lo que se puede aceptar que la diferencia sea de un factor 10, como se verá a continuación que es el caso.

Los 500 MW de potencia pretendidos (y declarados en la web oficial) en ITER son totales como energía cinética de los productos, es decir antes de su conversión en calor, mientras que los 3000 MW de un reactor de fisión típico son caloríficos. Ya se ha comentado que la energía total (la potencia no es más que energía por unidad de tiempo, luego se habla aquí de una y otra indiferentemente) producida en la fusión DT corresponde a energía cinética de sus productos, llevándose las partículas alfa aproximadamente el 20% y el restante 80% los neutrones rápidos. También se ha comentado que la transformación en calor de esta energía cinética se realiza mucho más eficientemente por parte de las partículas alfa (por estar cargadas eléctricamente) y es mucho más difícilmente aprovechable en el caso de los neutrones rápidos (que, en cambio, son mucho más eficientes activando radiactivamente los materiales de la vasija y produciendo daños estructurales, de ahí la necesidad de IFMIF-DONES). Siendo muy optimistas, podemos estimar que el 40% de la energía total liberada en foma de energía cinética de los productos en ITER (toda la de las alfas la cuarta parte de la de los neutrones rápidos) será transformable en calor[30]. Por lo tanto serían unos 200 MW en forma de calor los que previsiblemente se podrían extraer en ITER. Es decir, unas 10 veces menos que lo obtenido en un reactor de fisión *convencional*, lo cual es compatible con nuestras estimaciones.

Nos informa también la web oficial de ITER que se invertirán 50 MW de potencia entrante de calentamiento, pero no nos informa de la potencia eléctrica que será preciso suministrar para conseguir ese calor en el plasma. El factor Q utilizado es, como en el caso del NIF, un Q *científico*, cociente de la potencia total (no la calorífica, que es mucho menor) producida y la potencia útil para producir calentamiento (que es mucho menor que la total que se debe aportar). Como indiqué anteriormente, me parece una definición engañosa, ya que al menos habría que colocar en el numerador la potencia calorífica producida (es decir, la aprovechable, aunque parciamente), no la total, con lo cual el publicitado factor Q=500/50=10 se quedaría en Q=200/50=4. Además, al igual que en el caso del NIF, considero que en el denominador habría que colocar la potencia eléctrica total que se invertirá, que, como se ha mencionado, se estima entre 200 y 300 MW. De esta forma se obtiene un Q menor que la unidad, incluso antes de considerar la futura conversión en electricidad (delegada en DEMO), donde el Segundo Principio de la Termodinámica impondrá un rendimiento menor que la unidad al transformar la energía calorífica en energía eléctrica[31]. Con ello llegaríamos al Q *eléctrico*, que es el determinante: el cociente entre la potencia eléctrica vertida a la red (que habrá que esperar hasta su verificación en DEMO) y la potencia eléctrica consumida.

En mi opinión, las conclusiones que se extraen de lo anterior resultan evidentes. En todo caso, llama la atención que ITER cuente con una infraestructura capaz de aportar hasta 600 MW eléctricos gracias a la concentración de centrales nucleares de fisión en sus proximidades.

Es preciso resaltar que un cálculo riguroso de la potencia producida exige sofisticados cálculos y simulaciones por el método de Monte Carlo. Las anteriores estimaciones pretenden exclusivamente ilustrar sobre los órdenes de magnitud en que se desarrollan estos procesos, pero opinio que permiten hacerse una idea de conjunto, porque muchas veces, como suele decirse coloquialmente, "los árboles (de los innegables desarrollos científicos y tecnológicos) no dejan ver el bosque (de la finalidad última perseguida, que es la producción eficiente de energía eléctrica)".

**El tamaño importa**

Considero también pertinente mencionar el previsible tamaño de una central de fusión para producción de energía eléctrica, si algún día llega a construirse. Una de las muchas críticas que se han realizado en contra de las centrales nucleares de fisión es la gran concentración de infraestructuras y capital que implican y su tamaño, que van radicalmente en contra de una producción distribuida y cercana a los puntos de consumo. Ello sin olvidar los riesgos inherentes a dicha concentración provenientes de posibles ataques terroristas. A estas alturas del texto, creo que resulta evidente que en una central de fusión estos aspectos criticados en una central de fisión aumentan hasta dimensiones desconocidas hasta la fecha. No hay más que comparar el tamaño de ITER con su precursor JET y, aún más, con el previsto para DEMO. Los gabinetes de comunicación de los proyectos de fusión (NIF, ITER) nos inundan con informaciones grandilocuentes donde siempre aparece *lo más de lo más*: los láseres más potentes del mundo (en el caso del NIF), los imanes superconductores mayores del mundo, la vasija de vacío mayor del mundo, la soldadura electrónica más sofisticada del mundo, etc .. (en el caso de ITER). Son innegables logros de ingeniería a gran escala (y puede que ya eso de por sí justifique el esfuerzo y la energía invertidos), pero no deberían hacernos perder la visión de conjunto: de lo que se trata es de producir energía *aprovechable* en un futuro no demasiado lejano. Además, tampoco conviene olvidar que dicho desarrollo en busca de cuanto *más grande mejor* (porque esa es la única manera conocida de alcanzar las extremas condiciones descritas anteriormente) va en el sentido opuesto al seguido en los modernos prototipos de centrales de fisión modulares, destinados a su instalación a escala local, de los cual hay ya uno en fase operacional en Rusia[32] y otro en China en fase avanzada de construcción[33]; hay muchos otros diseños avanzados y prometedores en Japón, Europa y EEUU, que hasta ahora no se han podido llevar a la práctica. El hecho de que hayan sido precisamente Rusia y China los países que primero hayan llevado a la práctica esta idea innovadora, dice mucho del panorama geopolítico actual, donde la segunda (a Rusia aún le quedan rescoldos científicos y tecnológicos de la época soviética) se ha convertido a pasos agigantados en un referente mundial en ciencia y tecnología en todas las áreas estratégicas. Igualmente, se continúa investigando exhaustivamente en el ciclo de fisión del torio[34,35], desarrollando la tecnología para reactores más pequeños (llegando a la escala del MW), más seguros y con menos producción de residuos. No olvidemos que ITER , cuando entre en funcionamiento, consumirá del orden de 300 MW sólo para mantener la temperatura del plasma.

**Epílogo**

El proyecto ITER, al igual que la Estación Espacial Internacional (ISS, de sus siglas en inglés) surgieron en las mismas fechas (años 90) y con los mismos loables propósitos (fomentar la colaboración científico-técnica internacional), inmediatamente tras el derrumbe del bloque soviético y el comienzo de una época de absoluto dominio del bloque llamado occidental (aunque incluye también a Japón, Corea del Sur y, por supuesto, Australia y Nueva Zelanda) liderado por los EEUU de Norteamérica y la postración absoluta de la otra antigua potencia hegemónica, Rusia; China, aunque despegando, aún contaba poco. Era la época del famoso *Final de la Historia* de Francis Fukuyama. No creo necesario resaltar cómo ha cambiado el panorama internacional. La ISS, con Rusia retirándose, además de la poca relevancia de los resultados científicos obtenidos, está abocada a convertirse pronto en un trozo más de chatarra orbital destinada a desintegrarse en unos 10 años (si no antes, el silencio mediático es poco prometedor en ese sentido). Opino que, aparte de los innegables avances tecnológicos asociados a su desarrollo, ese es su principal ( y probablemente) único éxito.

   Los plazos han ido alargándose sin cesar: De la fecha inicialmente prevista de las primeras pruebas con plasma en ITER, 2016, se pasó a 2025 y hasta 2035 para las pruebas con la mezcla real DT. Los rumores sugieren insistentemente un nuevo alargamiento y la situación geopolítica mundial (al margen de los enormes problemas científico-técnicos asociados al proyecto) apunta a ello. Para DEMO ya ni siquiera se dan fechas concretas, sólo se habla de que será una realidad en la segunda mitad de la centuria. Vienen a la mente las palabras del Quijote: "¡Cuán largo me lo fiais, amigo Sancho!".

[1] Un nanosegundo es una milésima de una millonésima de segundo.

[2] El núcleo de hidrógeno, que consiste en un único protón, es el más simple de la Naturaleza y, obviamente es estable, es decir, no radiactivo. El de deuterio o deuterón consiste en un protón más un neutrón y es también estable. El de tritio consiste en un protón más dos neutrones y es radiactivo, con una semivida de 12.6 años (el tiempo en que tardan en desintegrarse la mitad de sus núcleos a partir de una población incial).

[3] Inicialmente la idea del *tokamak*, que es una palabra rusa porque fue un concepto propuesto en la Unión Soviética en los años 50, fue descartada por los investigadores norteamericanos, que llevaban desde los primeros años 50 trabajando en la Universidad de Princeton y el Laboratorio Nacional de Los Alamos tratando de controlar la fusión nuclear para usos civiles mediante confinamiento magnético con la configuración llamada *stellarator* (de aspecto exterior muy parecido al *tokamak*, pero con importantes diferencias conceptuales entre ambos), en paralelo al desarrollo del programa militar; de hecho ambos proyectos compartían inicialmente gran parte de su nombre en clave: Matterhorn-B para el militar (la B de *bomb*) y Matterhorn-S para la civil (con la S de *stellarator*) . A comienzos de los años 60 los científicos norteamericanos tuvieron que rendirse a la evidencia de que la idea del *tokamak* funcionaba mejor en muchos aspectos y comenzaron a investigar mayoritariamente basándose en ella, aunque varios grupos continuaron la investigación en la línea original (*stellarator*), que aún se mantiene hoy en día. Como puede fácilmente deducirse, la íntima interconexión en el entramado científico-militar norteamericano era evidente desde los inicios en estas investigaciones. Abundando en esta idea, cabe resaltar que el Livermore National Laboratory tiene una bien conocida vinculación íntima con la industria militar estadounidense en forma de contratos de investigación orientada tanto a las armas de fisión nuclear como a las basadas en láseres de alta potencia (sobre todo a partir de la llamada *Guerra de las Galaxias*, que promovió su presidente Ronald Reagan en los años 80). De hecho, la *I* del acrónimo NIF proviene de su objetivo inicial como centro de experimentación en explosivos nucleares mediante la compresión con láseres, sin necesidad de realizar explosiones, ya prohibidas por los tratados internacionales.

[4] La explicación del mecanismo de producción de energía en el Sol fue propuesta por Hans Bethe en su genial y clarividente artículo (uno más entre tantos de los suyos) "Energy Production in Stars", Physical Review 55 (1939) p. 434 , considerado uno de los diez mejores de la Historia de la Física moderna en una clasificación del Instituto Niels Bohr de Copenhague. Como curiosidad, este artículo fue inicialmente retirado por el autor para poder presentarlo a un concurso de ideas científicas inéditas (que obviamente ganó), con cuyo premio costeó la mudanza de su madre (judía perseguida en Alemania) a los Estados Unidos de América del Norte. En el mismo, entre otras muchas especulaciones basadas en las evidencias entonces disponibles, el autor propone una cadena que se inicia con la fusión de dos núcleos de hidrógeno, es decir dos protones, y que globalmente se traduce en que a partir de 4 protones se forma una partícula alfa, que es un núcleo de helio, con gran liberación de energía.

[5] A no ser que se consiga crear algún día en la Tierra algo parecido a una estrella de neutrones (en este caso de protones, es decir, un Sol .. pero necesitaríamos su masa para tener la compresión gravitacional suficiente, es decir, su tamaño, con lo cual nos quedaríamos sin Tierra ..); esto entra dentro del campo de la ciencia ficción.

[6] Es la combinación que presenta mayor probabilidad de fusión a las temperaturas que, aunque enormes (del orden de los cien millones de grados), pueden alcanzarse en una central de fusión.

[7] Adela Muñoz Páez, "Dónde está el litio", El Periódico, 28 de diciembre de 2022

[8] El término *termalización* significa alcanzar una distribución de velocidades correspondiente a esa temperatura de equilibrio, que, en Física, es la muy conocida distribución de Maxwell-Bolzmann. Mayor temperatura implica una mayor velocidad promedio a nivel microscópico, lo cual es fácilmente entendible a partir del concepto de temperatura

[9] [T. Giegerich et al, *Development of a viable route for lithium-6 supply of DEMO and future fusion power plants*, Fusion Engineering and Design, Volume 149, December 2019, 111339](#)

[10] En los experimentos realizados hasta la fecha en JET, la única instalación operativa por confinamiento magnético capaz de utilizarlo, se ha evitado en lo possible su utilización por las complicaciones que acarrea; de hecho, según la información de que dispongo, no se utiliza desde 1997.

[11] [Clery, Daniel (10 October 2013). "Fusion "Breakthrough" at NIF? Uh, Not Really …". Science.](#)

[12] 1 MeV es la unidad de energía típica de Física nuclear, siendo la energía cinética que adquiere un electrón acelerado por una diferencia de potencial de un millón de voltios.

13 Las partículas cargadas (en este caso las partículas alfa) son las responsables de la generación de la mayor parte del calor. El proceso es como sigue: al estar cargadas, interaccionan eléctricamente con los átomos del medio, arrancándoles electroles, es decir produciendo parejas iones positivos y electrones. Estos posteriormente se recombinan para formar nuevamente átomos neutros, liberándose la energía de ligadura correspondiente en forma de energías de vibración y rotación de dichos átomos, lo cual macroscópicamente se traduce en el aumento de temperatura. Los neutrones, por el contrario, son muy poco eficientes para producir calor a partir de su energía cinética (es decir aumento de temperatura del medio) debido a su carencia de carga eléctrica, que obliga a producir la ionización sólo indirectamente, a través de partículas cargadas secundarias.

14 1 MW es un millón de vatios. Para hacernos una idea: uno de los últimos reactores nucleares instalados en España produce del orden de 1000 MW de potencia eléctrica

15 Steven B. Krivit, "The ITER Power Amplification Myth". En New Energy Times, 6 Oct. 2017

16 Daniel Gassby, "ITER is a showcase … for the drawbacks of fusion energy", Bulletin of Atomic Scientists, February 14, 2018 .

17 https://www.iter.org/mach/Blanket

18 T. Giegerich et al, *Development of a viable route for lithium-6 supply of DEMO and future fusion power plants,* Fusion Engineering and Design, Volume 149, December 2019, 111339

19 Los primeros reactores nucleares de fisión estuvieron vinculados al programa militar norteamericano (proyecto *Manhattan*) . En concreto, la primera reacción en cadena controlada fue en el Pile-1 ("CP-1", acónimo de *Chicago Pile 1*) situado bajo el graderío oeste del campo de fútbol americano de la Universidad de Chicago, bajo la dirección científica de Enrico Fermi, el 2 de diciembre de 1942. Por ello se puede afirmar que la fisión nuclear controlada y la explosiva se desarrollaron en paralelo.

20 El nombre que inicialmente se le dio fue el de *bomba de hidrógeno*, o simplemente *bomba H* porque utilizaba isótopos de hidrógeno para producir la fusión, aunque el detonante fuera una bomba atómica (varias en los diseños modernos). Como dato se aporta el hecho de que aproximadamente las tres cuartas partes de la energía liberada en una explosión termonuclear proviene de la fisión, mientras que sólo la cuarta parte restante proviene de la fusión. Ello es debido a que un detonante central (*sparkplug,* o bujía, en su denominación inicial*)* de fisión *convencional* (Pu239 en el caso de Ivy Mike), que fisiona bajo el efecto de bombardeo con neutrones lentos, comprime y calienta la mezcla de isótopos de hidrógeno deuterio y tritio, que fusionan. En cada una de dichas fusiones se produce una partícula alfa (núcleo de helio) y un neutrón de alta energía que se utiliza para fisionar otro isótopo (U238 en Ivy Mike, dispuesto en la parte exterior del dispositivo), que precisamente fisiona muy eficientemente bajo el bombardeo con neutrones muy energéticos (no lo hace con neutrones lentos, por lo cual no es útil en las bombas atómicas convencionales) . Por lo tanto puede afirmarse que una bomba termonuclear, de hidrógeno o *de fusión* es en realidad un amplificador de la fisión mediante la fusión. De hecho, a pesar de su nombre, en realidad se trata de una bomba nuclear en 3 fases, fisión-fusión-fisión, por lo que más adecuadamente, en mi opinión, se la debería denominar *bomba A-H-A.* Esta disquisición semántica, que en apariencia puede parecer superflua, tiene una notable consecuencia en el imaginario colectivo, ya que se ha establecido la creencia de que en la fusión se libera mucha más energía que en la fisión.

21 La energía transportada por la radiación a temperaturas ordinarias es despreciable frente a la asociada a la agitación cinética a nivel microscópico, pero la primera aumenta con la cuarta ptencia de la temperatura, mientras que la segunada lo hace linealmente. A las enormes temperaturas alcanzadas en una reacción en una bomba de fisión la presión de la radiación se comporta como un gigantesco mazo que se utiliza para comprimir y calentar el plasma de fusión. Ello está descartado, obviamente, para aplicaciones pacíficas.

22 Un microsegundo es una millonésima de segundo.

23 La temperatura es una medida de la agitación a nivel microscópico, es decir de la energía cinética (aproximadamente proporcional a la velocidad al cuadrado en este contexto) de los constituyentes de la materia.

24 Número de partículas blanco por unidad de volumen. Obviamente en el caso de la fusión, proyectil y blanco son intercambiables, ya que ambos están en movimiento en el sistema del laboratorio, a diferencia de la fisión donde los blancos (usualmente núcleos de U235) están en reposo y son bombardeados por los neutrones.

25 El cálculo teórico y medida experimental de las secciones eficaces de las diferentes reacciones es, en última instancia, a lo que nos dedicamos los físicos nucleares.